\documentclass[twocolumn,showpacs,prl]{revtex4}

\usepackage{amsmath,subfigure,psfrag}
\usepackage{color}
\def\U#1{{\rm #1}} 
\def\u#1{_{\rm #1}}
\newcommand{\ket}[1]{| #1 \rangle}
\newcommand{\ketbra}[2]{| #1 \rangle \langle #2 |}
\newcommand{\expect}[1]{\langle #1 \rangle} 
\def\H{{\rm H}}
\def\V{{\rm V}}

\def\D{\U{D}}

\def\R{\U{R}}
\def\L{\U{L}}
\def\00{\H\V}
\def\11{\V\H}

\begin{document}
\title{
Efficient decoherence-free entanglement distribution 
over lossy quantum channels
}

\author{Rikizo Ikuta}
\affiliation{Graduate School of Engineering Science, Osaka University,
Toyonaka, Osaka 560-8531, Japan}
\author{Yohei Ono}
\affiliation{Graduate School of Engineering Science, Osaka University,
Toyonaka, Osaka 560-8531, Japan}
\author{Toshiyuki Tashima}
\affiliation{Graduate School of Engineering Science, Osaka University,
Toyonaka, Osaka 560-8531, Japan}
\author{Takashi Yamamoto}
\affiliation{Graduate School of Engineering Science, Osaka University,
Toyonaka, Osaka 560-8531, Japan}
\author{Masato Koashi}
\affiliation{Graduate School of Engineering Science, Osaka University,
Toyonaka, Osaka 560-8531, Japan}
\author{Nobuyuki Imoto}
\affiliation{Graduate School of Engineering Science, Osaka University,
Toyonaka, Osaka 560-8531, Japan}

\begin{abstract}
We propose and demonstrate a scheme 
for boosting up the efficiency of entanglement distribution 
based on a decoherence-free subspace (DFS) 
over lossy quantum channels. 
By using backward propagation of a coherent light, 
our scheme achieves 
an entanglement-sharing rate that is proportional 
to the transmittance $T$ of the quantum channel 
in spite of encoding qubits in multipartite systems for the DFS. 
We experimentally show that highly entangled states, 
which can violate the Clauser-Horne-Shimony-Holt inequality, 
are distributed at a rate proportional to $T$. 
\end{abstract}
\maketitle

Distribution of photonic entangled states among remote parties is 
an important issue in order to realize quantum information processing, 
such as quantum key distribution~\cite{QKD1, QKD2, QKD3}, 
quantum teleportation~\cite{teleportation}, 
and quantum computation~\cite{computation}. 
In practice, however, the quantum states are disturbed 
by fluctuations during the transmission. 
One of the possible schemes to overcome this problem 
is to encode the quantum states into a 
decoherence-free subspace~(DFS) in multipartite systems. 
In photonic systems, several proposals and 
experimental demonstrations have been done to 
show the robustness of quantum states 
in a DFS 
against collective fluctuations~\cite{DFS1,DFS2,DFS3,DFS4,DFS5,DFS6,DFS7}. 
Furthermore, the capability of faithful quantum-state 
transmission and entanglement distribution 
through an optical fiber have been demonstrated~\cite{cC,cY1,cY2}. 

A serious drawback of all photonic DFS schemes is that 
the photon losses in the quantum channel severely limit 
the transmission rate of quantum states, 
since all the photons forming the DFS must reach the receiver. 
When the quantum channel delivers a photon 
to the receiver with transmittance $T$, 
one can transmit a quantum state of interest 
only with a rate proportional to $T^n$ 
using an $n$-photon system with previous DFS 
schemes~\cite{DFS1,DFS2,DFS3,DFS4,DFS5,DFS6,DFS7,cC, cY1, cY2}. 
For realization of robust long distance quantum communication systems, 
it is thus desirable to improve the channel-transmission 
dependence of DFS schemes. 
In this Letter, 
we propose and experimentally demonstrate a 
two-photon DFS scheme 
for sharing entangled photon pairs, 
which boosts the efficiency to be proportional to $T$ 
from $T^2$ of the previous protocols in Ref.~\cite{cC,cY1,cY2}. 

\begin{figure}[t]
 \begin{center}
 \scalebox{1}{\includegraphics{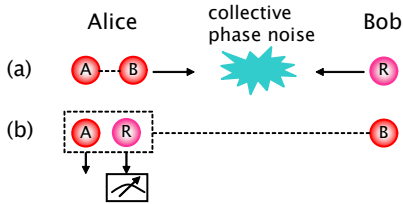}}
  \caption{(Color online) 
  Concept of our DFS scheme. 
  At step (a), Alice prepares a maximally entangled photon pair A and B, 
  and sends photon B to Bob's side. On the other hand, 
  Bob sends an ancillary photon R to Alice's side. 
  At step (b), Alice extracts the DFS 
  by the quantum parity checking on photons A and R, 
  and decodes back the initial entangled state from the DFS. 
  For boosting up the efficiency, 
  we use a coherent light instead of a single photon R.
  \label{fig:scheme}}
 \end{center}
\end{figure}

We first introduce our DFS scheme against collective phase fluctuations, 
as shown in Fig.~\ref{fig:scheme}. 
At step (a), 
the sender Alice generates 
a maximally entangled photon pair A and B 
in the state 
$\ket{\phi^+}\u{AB}
\equiv 
(\ket{\H}\u{A}\ket{\H}\u{B}+\ket{\V}\u{A}\ket{\V}\u{B})/\sqrt{2}$ 
and transmits photon B to Bob, 
where $\ket{\H}$ and $\ket{\V}$ represent 
horizontal~(H) and vertical~(V) polarization states of a photon, respectively. 
Meanwhile, 
the receiver Bob prepares an ancillary photon R 
in the state 
$\ket{\D}\u{R}\equiv (\ket{\H}\u{R}+\ket{\V}\u{R})/\sqrt{2}$, 
and sends photon R to Alice. 
After transmission of the photons, 
the states are transformed to 
$e^{i\phi\u{H}}\ket{\H\H}\u{AB}+e^{i\phi\u{V}}\ket{\V\V}\u{AB}$ and 
$e^{i\phi'\u{H}}\ket{\H}\u{R}+e^{i\phi'\u{V}}\ket{\V}\u{R}$ 
by the phase fluctuations in the channel, 
where $\phi\u{H(V)}$ and $\phi'\u{H(V)}$ represent 
phase shifts to the H~(V) components of photon B and R in the channel, 
respectively. 
Assuming that 
the difference between the phase shifts 
$\phi\u{H(V)}$ and $\phi'\u{H(V)}$ is negligibly small, 
the state of the three photons 
at the end of step (a) becomes 
\begin{widetext}
\begin{eqnarray}
\ket{\phi^+}\u{AB}\ket{\D}\u{R}
\rightarrow
\frac{1}{2}[
e^{i (\phi\u{H}+\phi\u{V})}
(\ket{\H}\u{B}\ket{\00}\u{AR}+\ket{\V}\u{B}\ket{\11}\u{AR})
+e^{2i \phi\u{H}}\ket{\H}\u{B}\ket{\H\H}\u{AR}
+e^{2i \phi\u{V}}\ket{\V}\u{B}\ket{\V\V}\u{AR}
]. 
\label{noise}
\end{eqnarray}
\end{widetext}
In this scheme, the pair of photons B and R 
that went through the collective noises 
end up being split between Alice and Bob. 
Nonetheless, Eq.~(\ref{noise}) can be interpreted 
as if the photons A and R, 
both of which are possessed by Alice, 
had gone through the collective noises. 
This comes from an important property 
of an entangled photon pair 
that a disturbance on one half of the photon pair is 
equivalent to a similar disturbance on 
the other half of the photon pair~\cite{Franson1, Franson2}. 
We find that the first two terms in Eq.~(\ref{noise}) 
are invariant under phase fluctuations. 
At step (b), 
by performing quantum parity checking 
on photons A and R~\cite{QPC}, 
Alice extracts the state in the DFS 
spanned by $\{ \ket{\00}\u{AR},\ket{\11}\u{AR} \}$ 
from the state in Eq.~(\ref{noise}). 
Then, 
the decoding back of the state into $\ket{\phi^+}\u{AB}$ is done 
by a projective measurement 
$\{ \ketbra{\U{D}}{\U{D}},\ketbra{\U{\bar{D}}}{\U{\bar{D}}} \}$ 
on photon R and a feedforward operation on photon A, 
where $\ket{\U{\bar{D}}}\equiv (\ket{\H}-\ket{\V})/\sqrt{2}$. 
When the transmittance of the quantum channel is $T$, 
the efficiency of this scheme is proportional to $T^2$ 
because both photons B and R must pass through the channel. 

Our strategy for enhancing the efficiency 
from $\mathcal{O}(T^2)$ to $\mathcal{O}(T)$ is 
to replace the single photon state in mode R by a coherent state. 
Suppose that the average photon number in mode R 
when received by Alice is $\mu$, 
namely, 
Bob initially prepares a coherent state of mean photon number 
$\mu\u{B}\equiv \mu T^{-1}$. 
With a probability of $\mathcal{O}(\mu T)$, 
Alice finds exactly one photon in mode R, 
and Bob also receives the photon B from Alice. 
The protocol then works exactly the same as was described before, 
leading to shared state $\ket{\phi^+}\u{AB}$. 
On the other hand, the use of the coherent state also 
produces unwanted events where two or more photons arrive at Alice, 
and a usual setup for quantum parity checking with linear optics and 
imperfect photon detectors cannot fully discriminate such events 
from the desired ones. 
Since these unwanted events occur with probability 
$\mathcal{O}(\mu^2 T)$, 
the condition $\mu \ll 1$ is needed to have 
a good fidelity of the final state. 
This condition is independent of $T$, 
which means that, given a target value of the fidelity, 
we may use a constant value 
of $\mu$~(and hence $\mu\u{B}$ proportional to $T^{-1}$) 
to reach the target for any value of $T$. 
This scheme thus gives a rate proportional to $T$ 
instead of $T^2$ in the previous two-photon DFS schemes. 

In our scheme, the counter-propagations of photons B and R are 
essential. If Bob prepares all the pulses A, B and R and sends A and R 
to Alice, 
the desired events occur 
with the same probability of $\mathcal{O}(\mu T)$ 
but the unwanted events occur 
with a larger probability of $\mathcal{O}(\mu^2)$, 
which makes the requirement on $\mu$ too stringent. 
Although the counter-propagation setup 
requires the phase fluctuations to be much 
slower than the propagation time, 
such a requirement has been experimentally shown 
to be met up to $\sim 100$ km 
in fiber-based quantum cryptography systems~\cite{PandP}. 

\begin{figure}[t]
 \begin{center}
 \scalebox{1}{\includegraphics{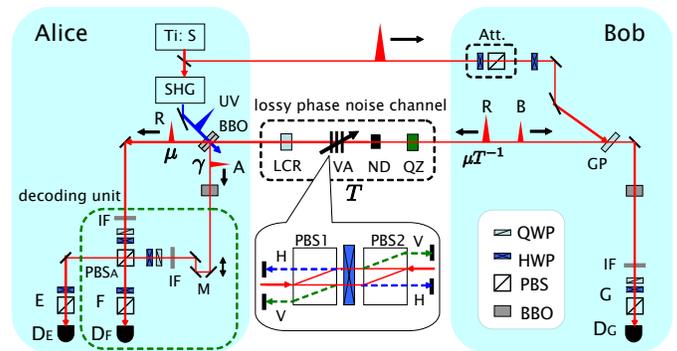}}
  \caption{(Color online) 
  Our experimental setup. 
  \label{fig:setup}}
 \end{center}
\end{figure}
The detail of our experimental setup is shown in Fig.~\ref{fig:setup}. 
We use a mode-locked Ti:sapphire~(Ti:S) laser~(wavelength: 790 nm; 
pulse width: 90 fs; repetition rate: 82 MHz) as a light source, 
which is divided into two beams. 
One beam is frequency doubled~(wavelength: 395 nm; power: 75 mW) 
by second harmonic generation~(SHG), 
and then pumps a pair of Type I phase-matched 1.5mm-thick 
$\beta$-barium borate~(BBO) crystals 
to prepare the entangled photon pair A and B 
through spontaneous parametric down conversion~(SPDC). 
The difference between the group velocities 
of H- and V-polarized photons is compensated by 
BBO crystals in each path of photon A and B. 
Photon A goes to Alice's decoding unit, 
while photon B enters a lossy phase noise channel 
and goes to the detector $\U{D}\u{G}$ 
after passing through a glass plate GP~(reflectance\ $\sim 5$\%). 
The other beam from the laser is used 
to prepare a coherent light pulse R at Bob's side. 
After adjusting the intensity of the coherent light pulse 
by an attenuator~(Att.) composed of a half-wave plate~(HWP) 
and a polarization beamsplitter~(PBS), 
we set its polarization to D by rotating a HWP. 
The coherent light pulse R is then reflected toward Alice's side by the GP, 
and enters the lossy phase noise channel. 
After that, it goes to Alice's decoding unit. 

To see the $T$ dependence of the rate, 
we use the lossy phase noise channel composed of 
a liquid crystal retarder~(LCR), 
a polarization-independent variable attenuator~(VA), 
a neutral-density filter~(ND) of transmittance $0.1$, 
and a quartz plate~(QZ). 
The LCR provides a phase shift 
between $\ket{\H}$ and $\ket{\V}$ 
according to the applied voltage. 
For simulating the collective random phase fluctuations, 
we slowly switched among eight values of phase shifts, 
$n\pi/4~(n=0,\ldots ,7)$, 
such that pulses A and R undergo the same fluctuations. 
This simulates 
the cases where the phase fluctuations are much slower 
than the propagation time of pulses A and R 
in fiber-optic communication. 
The VA is composed of 
a HWP sandwiched with two calcite PBSs, 
which deflect V-polarized photons, 
as shown in a subfigure inserted in Fig.~\ref{fig:setup}. 
By rotating the HWP, we can vary transmission $T$ continuously 
for both polarizations. 
The QZ compensates an additional group delay introduced by PBS1 and PBS2. 
In addition to the variable loss, 
the VA also swaps the H- and V-polarization components 
of light. 
This effect would be removed by inserting another HWP, 
but in our experiment, we cancel it 
by a proper re-labeling of polarizations in Bob's apparatus. 

Alice's detection unit carries out 
quantum parity checking for extracting the DFS 
and decoding for recovering the entangled state~\cite{QPC}, 
when each of modes A and R has a single photon. 
After receiving the pulse R, 
Alice inverts its polarization by a HWP 
before $\U{PBS}\u{A}$. 
Adjusting a temporal delay by mirrors~(M) 
on a motorized stage, 
Alice mixes the pulses A and R at $\U{PBS}\u{A}$, 
and post-selects later the cases where 
there is at least one photon in each mode E and F. 
This operation is the quantum parity checking, 
which discards the cases 
where the input state of photons A and R was 
$\ket{\H\H}\u{AR}$ or $\ket{\V\V}\u{AR}$. 
In mode F, Alice selects the cases where 
the photon is projected onto $\ket{\U{D}}\u{F}$ 
by the detector $\U{D}\u{F}$ with a HWP and a PBS. 
The final state of the shared photon pair E and G, 
which should be $\ket{\phi^+}\u{EG}$ ideally, 
is then analyzed by projecting the photons E and G 
to various polarizations, H, V, D, and $\U{\bar{D}}$. 
The collected data is thus composed of 
the rates of triple coincidence events 
among $\U{D}\u{E}$, $\U{D}\u{F}$ and $\U{D}\u{G}$ 
for different rotation angles of HWPs in front of 
$\D\u{E}$ and $\D\u{G}$. 
The spectral filtering of the photons for all detectors is performed 
by narrow-band interference filters~(IF, wavelength: 790 nm; 
bandwidth: 2.7 nm). 
All the detectors $\D\u{E}$, $\D\u{F}$ and $\D\u{G}$ 
are silicon avalanche photodiodes 
which receive photons through single-mode optical fibers. 

In our experiments, 
we use SPDC 
with a photon pair generation rate $\gamma$ 
as the entangled photon source. 
In this case, 
an additional condition 
between $\mu$ and $\gamma$ is required 
to reduce false triple coincidences 
caused by the multiple photon pair generation from SPDC, 
whose probability is $\mathcal{O}(\gamma^2 T)$. 
Since the true coincidences occur 
at probability $\mathcal{O}(\mu \gamma T)$, 
the condition $\gamma\ll \mu$ is required. 
Therefore, 
to achieve a high fidelity, 
we need to satisfy $\gamma \ll \mu \ll 1$. 
In the following experiments, 
we set 
$\gamma\approx 3.0\times 10^{-3}$ 
and $\mu\approx 1.1\times 10^{-1}$. 

\begin{figure}[t]
 \begin{center}
 \scalebox{1}{\includegraphics{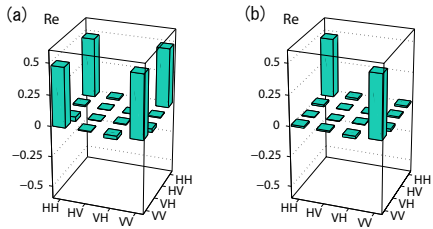}}
  \caption{(Color online) 
  The real parts of 
  (a) $\rho\u{AB}$, and (b) $\rho'\u{AB}$. 
  \label{fig:matrix}}
 \end{center}
\end{figure}
As preliminary experiments, 
we characterized the two-photon state from SPDC 
by recording the coincidence events 
between $\U{D}\u{E}$ and $\U{D}\u{G}$ 
without sending the photons in mode R. 
We chose $T=0.1$ and performed quantum state tomography 
by rotating the quarter-wave plate~(QWP) 
and the HWP before $\U{PBS}\u{A}$ at Alice's side, 
and rotating the QWP and the HWP in mode G at Bob's side~\cite{MQ}. 
Without the phase fluctuations, 
the density operator $\rho\u{AB}$ 
of the two-photon state is reconstructed as in Fig.~\ref{fig:matrix}~(a). 
The iterative maximum likelihood method was used 
for the reconstruction~\cite{MLE1, MLE2}. 
The observed fidelity of $\rho\u{AB}$ 
to the maximally entangled state $\ket{\phi^+}\u{AB}$ 
was $0.98\pm 0.01$, 
which implies the photon pair prepared by Alice 
was in a highly entangled state. 
Fig.~\ref{fig:matrix}~(b) shows the state $\rho'\u{AB}$ 
with the phase fluctuations. 
We see that the off-diagonal elements vanished as expected, 
indicating that the phase noises by the LCR 
effectively simulated the random phase noise channel. 
The observed fidelity of $\rho'\u{AB}$ was $0.51\pm 0.01$. 

\begin{table}
\begin{center}
\begin{tabular}
{cccc}
\hline
$T$ & $V\u{Z}$ & $V\u{X}$ & $F\u{low}$\\ \hline
0.1 & $0.88\pm 0.02$ & $0.82\pm 0.03$ & $0.85\pm 0.02$\\
0.03 & $0.91\pm 0.02$ & $0.79\pm 0.03$ & $0.85\pm 0.02$\\
0.01 & $0.88\pm 0.02$ & $0.77\pm 0.03$ & $0.82\pm 0.02$\\
0.005 & $0.82\pm 0.03$ & $0.72\pm 0.04$ &$0.77\pm 0.02$\\
0.003 & $0.74\pm 0.03$ & $0.66\pm 0.04$ & $0.70\pm 0.03$\\ \hline
\end{tabular}
 \caption{
 The observed visibilities ($V\u{Z}$ and $V\u{X}$) and 
 a lower bound $F\u{low}=(V\u{Z}+V\u{X})/2$ on the shared state 
 for channel transmittance $T$. 
  \label{tbl:view}}
 \end{center}
\end{table}
\begin{figure}[t]
 \begin{center}
 \scalebox{1}{\includegraphics{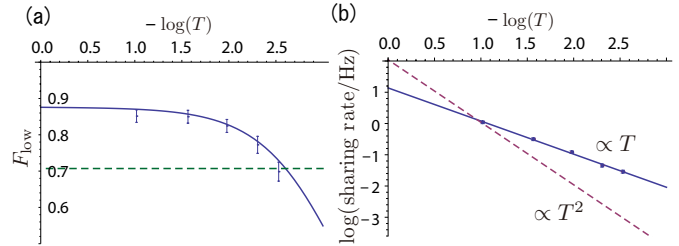}}
  \caption{(Color online) 
  (a) The dependence of $F\u{low}$ on the transmittance $T$ in dB. 
  Dots with error bars are the derived values of $F\u{low}$ 
  from $V\u{X}$ and $V\u{Z}$ in Table~\ref{tbl:view}. 
  The solid curve is obtained by theoretical calculation with 
  $V\u{sp}\approx 0.90$ and experimental parameters. 
  The broken line indicates the lower bound of the fidelity 
  to a maximally entangled state 
  to see the violation of the CHSH inequality. 
  (b) The experimental sharing rate of output states. 
  The slope of the solid line fitted 
  to the experimental data is $1.06\pm 0.04$, 
  which clearly shows the sharing rate is proportional to $T$. 
  The broken line describes where an ideal single photon 
  is used for mode R instead of the coherent light pulse, 
  whose rate is proportional to $T^2$. 
  The two lines are expected to intersect at $T=\mu$. 
  \label{fig:result}}
 \end{center}
\end{figure}
We then performed our DFS scheme. 
The quality of the shared entangled state 
was evaluated 
by determining two visibilities 
$V\u{Z}\equiv \expect{Z\u{E}Z\u{G}}$ and 
$V\u{X}\equiv \expect{X\u{E}X\u{G}}$ 
from the observed coincidence rates, 
where 
$Z\equiv \ketbra{\H}{\H}-\ketbra{\V}{\V}$ and 
$X\equiv \ketbra{\D}{\D}-\ketbra{\U{\bar{D}}}{\U{\bar{D}}}$. 
A lower bound $F\u{low}$ of the fidelity is then given by 
$F\u{low}=(V\u{Z}+V\u{X})/2$~\cite{fidelity}. 
$F\u{low}>1/\sqrt{2}\sim 0.707$ implies that 
the observed two photons are strongly entangled 
and can violate the Clauser-Horne-Shimony-Holt~(CHSH) inequality. 
The experimental value of $F\u{low}$ at $T=0.1$ 
was $0.85\pm 0.02$, 
which shows 
our DFS scheme well protects 
the quantum correlations against phase fluctuations. 
Next, we demonstrated our DFS scheme for various values of $T$ 
ranging from $0.1$ to $0.003$. 
We chose the intensity of the coherent light pulse R 
at Bob's side to be proportional to $T^{-1}$, 
such that $\mu$ be a constant. 
Table~\ref{tbl:view} shows the results 
of observed visibilities 
and the derived values of $F\u{low}$. 
Visibility $V\u{Z}$ is generally better than $V\u{X}$ 
since only the latter is affected by mode mismatch  
between pulses A and R at $\U{PBS}\u{A}$. 
We plot the relationship between $T$ and $F\u{low}$ 
in Fig.~\ref{fig:result}~(a), 
which implies that 
the shared states between Alice and Bob were highly 
entangled for $T\geq 0.005$. 
The sharing rate of output states at each $T$ is 
shown in Fig.~\ref{fig:result}~(b). 
We clearly see that the sharing rate is proportional to $T$. 
A broken line in Fig.~\ref{fig:result}~(b), 
which is proportional to $T^2$, is the rate 
expected when Bob uses an ideal single photon for mode R. 
By comparison, we see that our scheme is favorable 
for smaller values of $T$ as long as 
the observed values of $F\u{low}$ are acceptable. 

In order to see the reason of 
the degradation of $F\u{low}$ for small $T$, 
we constructed a simple theoretical model 
which regards each pulse as a single mode 
but takes into account multi-photon emission events 
and the mode matching $V\u{sp}$ between modes A and R. 
We used the following experimental parameters in the model: 
$\gamma\approx 3.0\times 10^{-3}$, 
$\mu\eta\approx 1.4\times 10^{-2}$, 
$\eta \approx 0.13$, 
$\eta\u{G} \approx 0.09$ 
and $d\approx 1.5\times 10^{-6}$. 
Here, $\eta$ is the quantum efficiency of $\U{D}\u{E}$ and $\U{D}\u{F}$, 
$\eta\u{G}$ is the quantum efficiency of $\U{D}\u{G}$, 
and $d$ is the dark count rate of $\U{D}\u{G}$. 
The value of $V\u{sp}$ was then determined to be $0.90$ 
by requiring that the model should correctly 
predict the observed value of $V\u{X}= 0.82$ at $T=0.1$. 
With no other adjustable parameters, 
the theory predicts the solid curve in Fig.~\ref{fig:result}~(a), 
which is in good agreement with the observed values. 
In the theoretical model, 
the degradation of $F\u{low}$ for small $T$ is mainly 
caused by the relative increase of the contribution 
from the dark counts of Bob's detector~$\U{D\u{G}}$. 
Hence, 
the degradation of the fidelity will be avoided 
by using low dark-count detectors, 
such as 
superconducting single-photon detectors used 
in QKD experiments~\cite{SSPD}. 

\begin{figure}[t]
 \begin{center}
 \scalebox{1}{\includegraphics{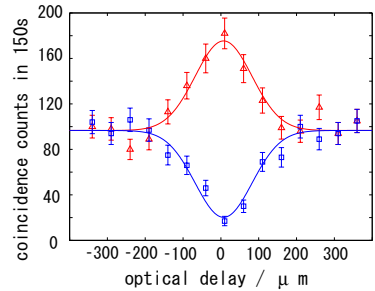}}
  \caption{(Color online)
  The observed quantum interference 
  by mixing photon A in the state $\ket{\R}\u{A}$ 
  and the coherent light pulse R with D polarization, 
  where $\ket{\R}\equiv \ket{\H}+i\ket{\V}$. 
  The triangles and squares show the coincidence counts 
  measured on the bases $\ket{\R}\u{E}\ket{\U{D}}\u{F}$ 
  and $\ket{\L}\u{E}\ket{\U{D}}\u{F}$, respectively. 
  Here $\ket{\L}\equiv \ket{\H}-i\ket{\V}$. 
  The visibility at the zero delay is $0.83\pm 0.04$. 
  \label{fig:visibility}}
 \end{center}
\end{figure}
We note that the interference occuring at $\U{PBS}\u{A}$ 
is robust against timing mismatch 
between photon A and the coherent light pulse R. 
Fig.~\ref{fig:visibility} shows 
observed quantum interference 
as a function of the optical delay 
introduced by moving the mirrors M in Fig.~\ref{fig:setup}. 
The FWHM is calculated as $\sim 180\mu\U{m}$. 
This value is over 200 times larger 
than the photon wavelength $790\U{nm}$, 
which implies that wavelength-order precision 
of the control is not required in our scheme. 
While we have derived the coherent light pulse of Bob 
from the pump laser sitting on Alice's side 
for simplicity of our experiment, 
the robustness against timing fluctuations suggests 
that the coherent light pulse can be independently prepared by Bob. 
Such two photon interference experiments 
using independently prepared pump lasers 
have been demonstrated in Ref.~\cite{sync1, sync2}. 

We have proposed and demonstrated 
an efficient decoherence-free entanglement-sharing scheme 
with a rate proportional to the transmittance of the quantum channel. 
In our scheme, 
the property of an entangled photon pair, 
that a phase disturbance on one half can be 
cancelled at the other side, 
enables us to use counter-propagations of the two photons. 
This permits us to 
use a coherent light pulse 
with the prepared intensity inversely proportional to the transmittance 
of the channel as an ancillary system, 
which leads to boosting up of the efficiency of 
entanglement distribution. 
Because the phase-cancellation property holds true 
for any state of the form $\alpha \ket{\H\H}+\beta\ket{\V\V}$, 
our DFS scheme is applicable to 
distribution of any unknown single qubit $\alpha \ket{\H}+\beta\ket{\V}$ 
by encoding it into 
$\alpha \ket{\H\H}+\beta\ket{\V\V}$ 
using quantum parity checking~\cite{QPC}. 
We believe that the proposed scheme is useful 
for realizing stable long-distance quantum communication~\cite{DLCZ}. 

We thank Tsuyoshi Kitano and \c{S}ahin K. \"Ozdemir 
for helpful discussions. 
This work was supported by the Funding Program for 
World-Leading Innovative R \& D on Science and 
Technology (FIRST), MEXT Grant-in-Aid for Scientific 
Research on Innovative Areas  20104003 and 21102008, 
JSPS Grant-in-Aid for Scientific Research (C) 20540389, 
and the MEXT Global COE Program.


\begin{thebibliography}{999}

\bibitem{QKD1} C. H. Bennett and G. Brassard, 
	in Proceedings of IEEE International Conference on Computers, 
	Systems, and Signal Processing, 
	Bangalore, India (IEEE, New York, 1984), p. 175.  
\bibitem{QKD2} A. K. Ekert, 
	\prl {\bf 67}, 661 (1991).
\bibitem{QKD3} C. H. Bennett, G. Brassard, and N. D. Mermin, 
	\prl {\bf 68}, 557 (1992).
\bibitem{teleportation} C. H. Bennett {\it et al.}, 
	\prl {\bf 70}, 1895 (1993). 
\bibitem{computation}
	R. Raussendorf and H. J. Briegel, 
	\prl {\bf 86}, 5188 (2001).
\bibitem{DFS1} P. G. Kwiat {\it et al.}, 
	Science, {\bf 290}, 498 (2000). 
\bibitem{DFS2} Z. D. Walton {\it et al.}, 
	\prl {\bf 91}, 087901 (2003).
\bibitem{DFS3} J.-C. Boileau {\it et al.}, 
	\prl {\bf 92}, 017901 (2004).
\bibitem{DFS4} M. Bourennane {\it et al.}, 
	\prl {\bf 92}, 107901 (2004).
\bibitem{DFS5} J.-C. Boileau {\it et al.}, 
	\prl {\bf 93}, 220501 (2004).
\bibitem{DFS6} T. Yamamoto {\it et al.}, 
	\prl {\bf 95}, 040503 (2005).
\bibitem{DFS7} R. Prevedel {\it et al.}, 
	\prl {\bf 99}, 250503 (2007).
\bibitem{cC} T.-Y. Chen {\it et al.}, 
	Phys. Rev. Lett. {\bf 96}, 150504 (2006). 
\bibitem{cY1} T. Yamamoto {\it et al.}, 
	New J. Phys. {\bf 9}, 191 (2007). 
\bibitem{cY2} T. Yamamoto {\it et al.}, 
	Nature Photonics {\bf 2}, 488 (2008). 
\bibitem{PandP} D. Stucki {\it et al.}, 
	New J. Phys. {\bf 4}, 41 (2002). 
\bibitem{Franson1} J. D. Franson, \pra {\bf 45}, 3126 (1992). 
\bibitem{Franson2} J. D. Franson, \pra {\bf 80}, 032119 (2009). 
\bibitem{QPC} T. B. Pittman, B. C. Jacobs, and J. D. Franson, 
	\pra {\bf 64}, 062311 (2001). 
\bibitem{MQ} D. F. V. James {\it et al.}, 
	\pra {\bf 64}, 052312 (2001). 
\bibitem{MLE1} J. $\U{\check{R}eh\acute{a}\check{c}ek}$ {\it et al.}, 
	\pra {\bf 75}, 042108 (2007). 
\bibitem{MLE2} T. Tashima {\it et al.}, 
	\prl {\bf 102}, 130502 (2009). 
\bibitem{fidelity} K. Nagata, M. Koashi, and N. Imoto, 
	\pra {\bf 65}, 042314 (2002). 
\bibitem{SSPD} S. Miki {\it et al}, 
	Opt. Lett., {\bf 35}, 2133, (2010). 
\bibitem{sync1} R. Kaltenbaek {\it et al.}, 
	\prl {\bf 96}, 240502 (2006). 
\bibitem{sync2} T. Yang {\it et al.}, 
	\prl {\bf 96}, 110501 (2006). 
\bibitem{DLCZ} L.-M. Duan {\it et al.}, 
	Nature {\bf 414}, 413 (2001). 
\end{thebibliography}
\end{document}